\documentclass[12pt]{iopart}

\def\snn{\mbox{$\sqrt{s_{_{\rm NN}}}$}}

\newcommand{ \be }{\begin{eqnarray}}
\newcommand{ \ee }{\end{eqnarray}}
\newcommand{ \la }{\langle}
\newcommand{ \ra }{\rangle}

\newcommand{ \bp }{{\bf p}}
\newcommand{ \bP }{{\bf P}}
\newcommand{ \bV }{{\bf V}}
\def \br {{\bf r}}
\newcommand{ \bq }{{\bf q}}

\newcommand{ \mean }[1]{\la #1 \ra}

\usepackage{graphicx}
\begin{document}

\title[Femtoscopy of the system shape fluctuations]
{Femtoscopy of the system shape fluctuations in heavy ion collisions}

\author{Sergei A. Voloshin}
\address{Wayne State University,
666 W. Hancock,
Detroit, Michigan 40201}
\ead{voloshin@wayne.edu}

\begin{abstract}
Dipole, triangular, and higher harmonic
flow that have an origin in the initial density fluctuations
has gained a lot of attention as they can provide additional
important information about the dynamical
properties (e.g. viscosity) of the system.
The fluctuations in the initial geometry should be also reflected in the
detail shape and velocity field of the system at freeze-out. 
In this talk I discuss 
the possibility to measure such fluctuations 
 by means of identical and non-identical particle interferometry.  
\end{abstract}



The initial spatial conditions in heavy ion collisions are not azimuthally
symmetric. Due to particle rescatterings it leads to
the anisotropic flow -- anisotropies in particle momentum
distributions -- with elliptic flow being the well known example. 
Recently, a
significant progress has been reached in understanding the role of
the initial density fluctuations. In particular it was realized
that such fluctuations lead to odd harmonic anisotropic
flow, studying of which brings new insights into dynamics of the
system evolution. Below, I briefly
summarize recent developments. I apologize for not providing 
the full list of references -- many of which can be found in other proceedings of this
conference. 

An importance of the initial state fluctuations for flow development
has been noticed in~\cite{Aguiar:2001ac}, though the exact relation
between fluctuation and the final event anisotropies were not clear.
In~\cite{Voloshin:2003ud} an observation was made that non-zero
two-particle rapidity correlations (as observed in $pp$ collisions) in
conjunction with radial flow (accepted part of the system evolution in
$AA$ collisions) lead to a narrow in azimuth and long ranged in
rapidity correlations. Such correlations were observed
experimentally and called {\em ridge}.  This mechanism (non-zero
``primordial'' rapidity correlations + radial flow $\Rightarrow$
ridge) have been exploited later in several other models that try to
address in more detail a (more subtle) question of the origin of the
``primordial'' rapidity correlations.  Although the appearance of the
ridge and anisotropic flow fluctuations look as totally unrelated
phenomena, they appeared to be different views on the same thing --
reaction of the system to the fluctuation in the initial density.
In~\cite{Takahashi:2009na} it was shown explicitly that the
fluctuations in the initial density distribution that extends over
large rapidity range lead to the ridge structure in two particle
correlations.  It was also noticed that after subtraction of the
expected contribution from elliptic flow, two-bump structure appears
on the away-side.  Further studies~\cite{Andrade:2010sd} with a single
''hot spot'' embedded into otherwise smooth background appeared quite
revealing. Contrary to an expectation~\cite{Voloshin:2003ud} that such
a hot spot moved out by the radial flow would create a ``bump'' in
azimuthal distribution it appeared that that the hot spot actually
``blocks'' the development of the radial flow and leads to a ``dip''
in particle distribution at the corresponding azimuth, accompanied by
side-splashes from both sides.  Remarkably, in terms of the
correlation function the ``bump'' and the ``dip'' leads to the same
structure - the ridge, as both means {\em positive} correlations. At
the same time, the exact shape of the correlation function (e.g. the
strength of the third harmonic, which is important below) is
different.

It was noticed in~\cite{Mishra:2008dm,Sorensen:2010zq} that the
fluctuating initial conditions generate anisotropic flow of different
harmonics.  That followed by
understanding~\cite{Teaney:2010vd,Staig:2011wj,Shuryak:2009cy} that
the perturbations due to each of the hot spots can be treated
independently, which allows to reformulate the problem from a
different perspective -- decompose the initial density into multipoles
and study the system responce to each of the multipoles - the approach
widely discussed at this conference as flow  (dipole,
triangular, quadrangular, etc.)  fluctuations.
One can envision such an
approach as rotating each of the event to a given harmonic symmetry
plane, such that at the end one could have a smooth initial conditions
but with a shape corresponding to a given harmonic.  The question
addressed in this talk is if with the help of femtoscopy
(identical and non-identical two-particle correlations) one can
directly observe such triangular etc. shapes of the system.  In this study I
follow an approach of the first papers where the azimuthally
sensitive femtoscopic analyses have been
proposed~\cite{Voloshin:1995mc,Voloshin:1997jh}, and 
blast wave and AMPT models calculations.

\begin{figure}[h!]
\includegraphics[width=55mm]{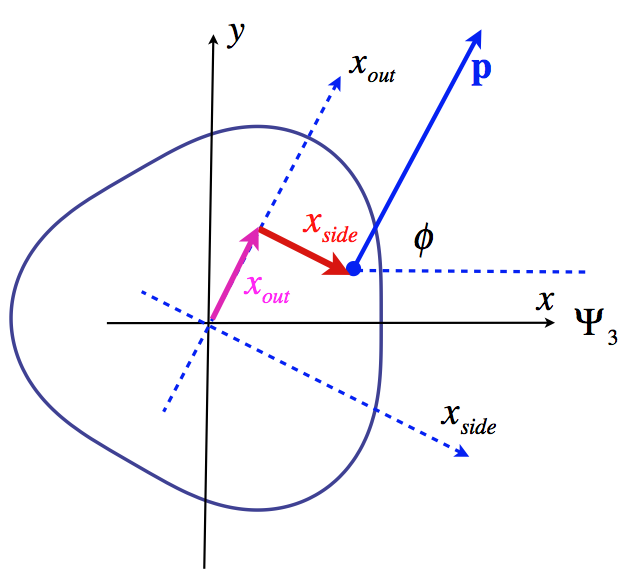}
\hspace*{-1.0cm}
\includegraphics[width=51mm]{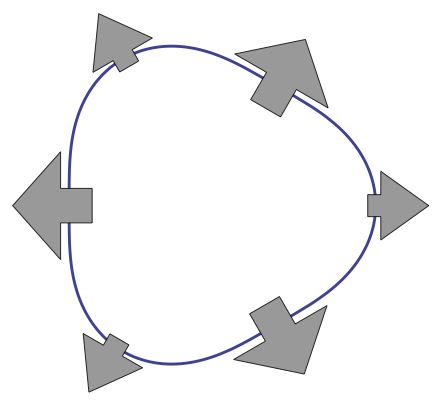}
\includegraphics[width=60mm]{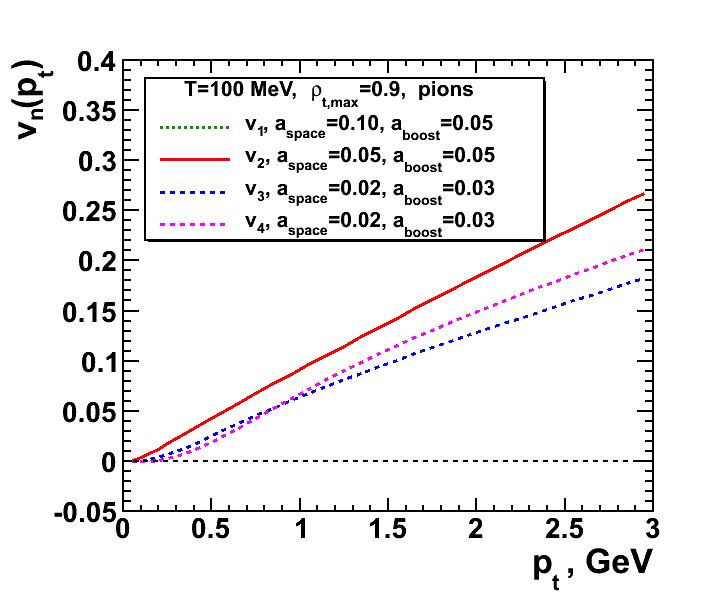}
\caption{Left: side-out coordinates. Middle: illustration
  of predominant expansion along shorter directions.
$v_n(pt)$ for typical values of parameters used in this work. 
}
\label{fig:coordinates} 
\end{figure}

Intensity interferometry (identical particle femtoscopy), measures
particle space-time distribution at freeze-out. Due to the probe on-mass
constraint, the
full space-time information can not be obtained, and one can access 
only the distribution in  $(\br-\bV t)$, where $\br$ is the
particle position at time $t$ (after freeze-out), and
$\bV$ is the particle velocity. For a Gaussian source, the correlation
function appears to be a Gaussian 
\be
C(\bq, \bP) \propto 1+ \exp \sum_{i,j} R_{ij}^2 q_i q_j,
\ee
where $\bq=\bp_1-\bp_2$ is the pair relative momentum,
 $\bP=(\bp_1+\bp_2)/2\approx \bp_i$ is the particle average momentum,
 and $R_{ij}^2=\mean{(r_i-V_it)(r_j-V_jt)}$ are the HBT radii. 
The deviation from a Gaussian form can be studied by evaluating the higher
moments of the distribution (for the effect of non-Gaussiness on HBT
radii, see~\cite{Hardtke:1999vf}). Detailes of femtoscopic analyses
can be found in a review~\cite{Lisa:2005dd}.

\begin{figure}[h!]
\begin{minipage}[t]{168mm}
\includegraphics[width=58mm]{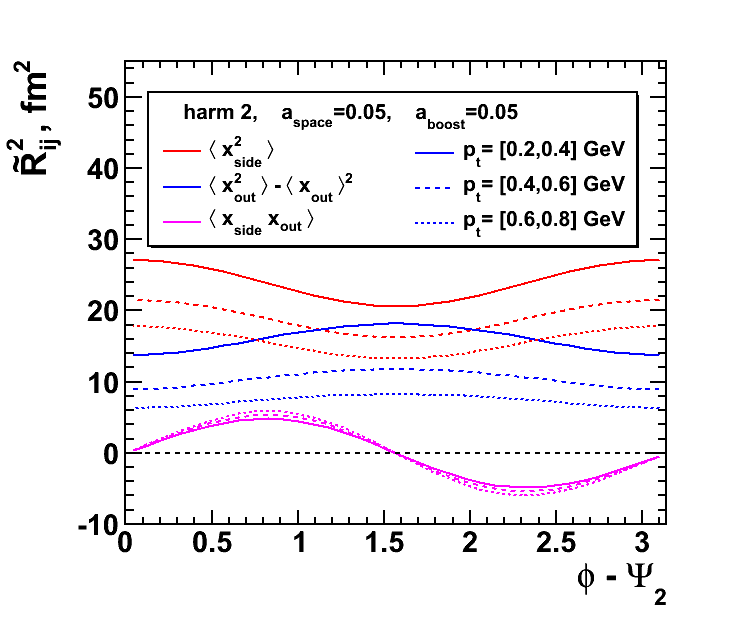}
\hspace*{-0.7cm}
\includegraphics[width=58mm]{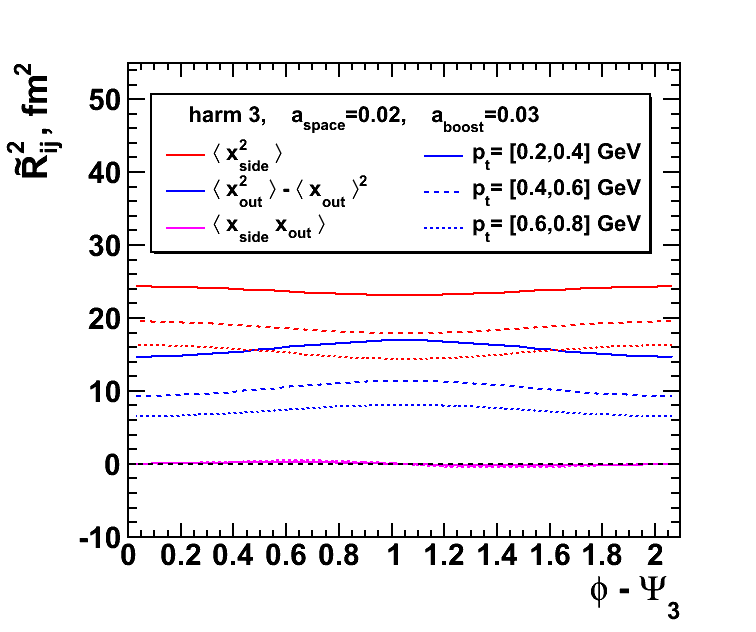}
\hspace*{-0.7cm}
\includegraphics[width=58mm]{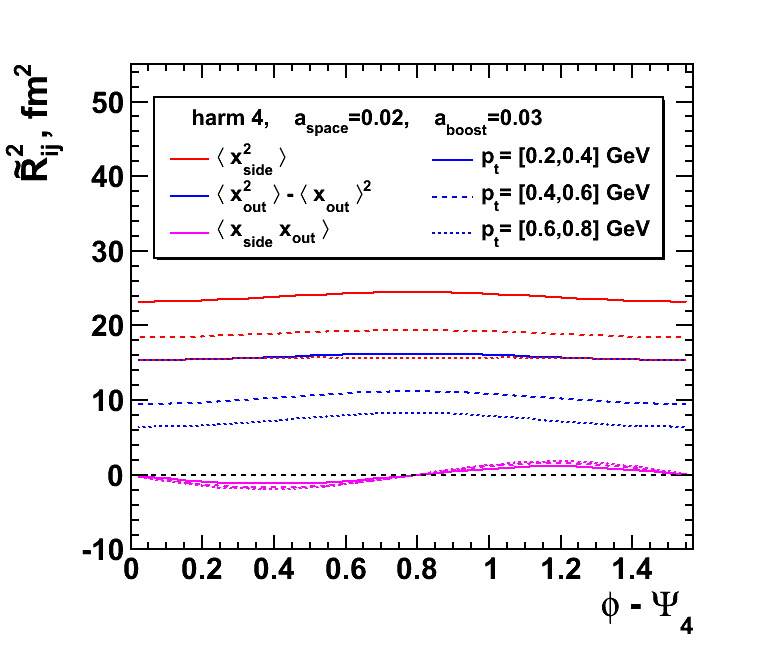}
\end{minipage} 
\caption{Blast wave model: HBT radii azimuthal dependence relative to
  the corresponding symmetry planes for harmonics 2, 3, and 4.}
\label{fig:BWradii}
\end{figure}

We use a standard side-out-long system~\cite{Lisa:2005dd} (see
figure~\ref{fig:coordinates}) and concentrate on azimuthal dependence
of the side, out, and side-out radii.  For a {\em stationary} (not
expanding) source the radii azimuthal dependence can be expressed as
\be
R_{side}^2=\mean{x_{side}^2} = \mean{x^2}\sin^2\phi
+\mean{y^2}\cos^2\phi-\mean{xy}\sin2\phi, 
\ee 
which has only $n=2$ harmonic.  Higher harmonics azimuthal dependence
appears only as deviation from the Gaussian, e.g. in
$\mean{x_{side}^6}$ and $\mean{x_{side}^4}$ for triangular and
quadrangular shapes respectively.  The picture changes if one
considers azimuthal variation in the expansion velocity. To study this
effect we employ a blast wave model.  We assume longitudinally boost
invariant source and freeze-out at a constant temperature
$T=100$~MeV. Expansion velocity profile is parametrized in a form
\be
\rho_t(r,\phi)=\rho_{t,max}\frac{r}{r_{max}(\phi)}[1+a_{boost}
  \cos(n\phi)], \\ r_{max}=R[1+a_{space} \cos(n\phi)], 
\ee 
with direction of the collective velocity being perpendicular to the
line $\propto [1+a_{space} \cos(n\phi)]$ going through the emission
point.  The parameter $\rho_{t,max}=0.9$ which corresponds to an
average expansion velocity $\mean{v_r}\approx 0.7$ and
$\mean{p_t}_{\pi}=0.40$~GeV.  $a_{space}$ and $a_{boost}$ parameters
define the spatial eccentricity at freeze-out and modulation in the
expansion velocity.  For the first we assume that the system spatial
eccentricity is reduced about two times during the system evolution,
which leads to $a_{space} \sim 0.05$; $a_{boost}$ can be estimated
from $v_n(p_t)$ dependence and is taken to be in the range 0.03-0.07.
The azimuthal dependence of the obtained HBT radii, see
figure~\ref{fig:BWradii}, indicates that the higher harmonics
shape effects become clearly visible (and measurable). Note the
difference in the amplitudes of $R_{side-out}$ terms. The radii
modulation strongly depends on the parameters of the model, and the
correspondng measurements will be very important in evaluating the
velocity profiles at freeze-out and testing models.

\begin{figure}[h!]
\begin{minipage}[t]{168mm}
\includegraphics[width=58mm]{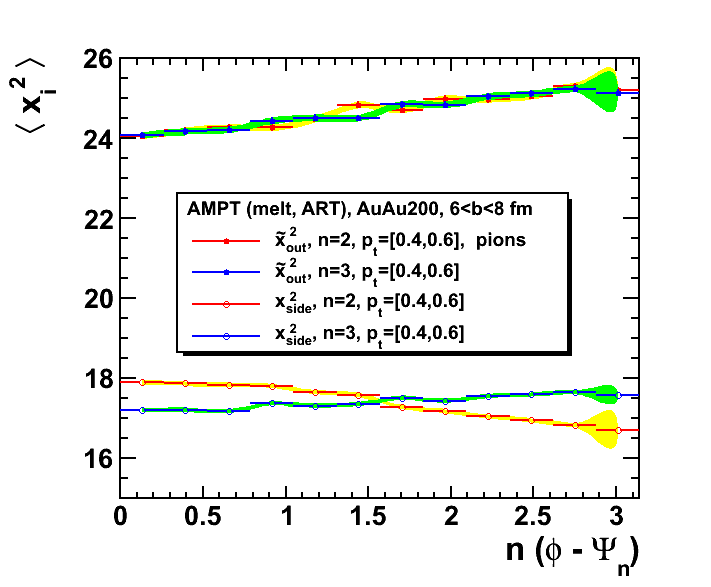}
\hspace*{-0.71cm}
\includegraphics[width=58mm]{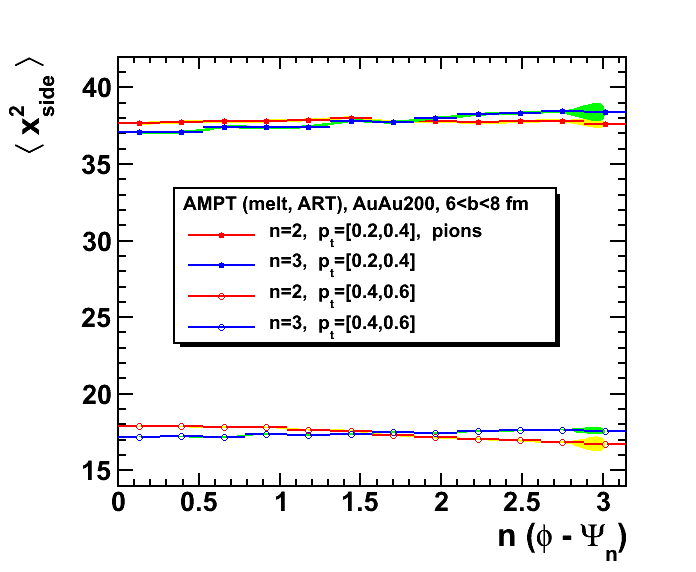}
\hspace*{-0.71cm}
\includegraphics[width=58mm]{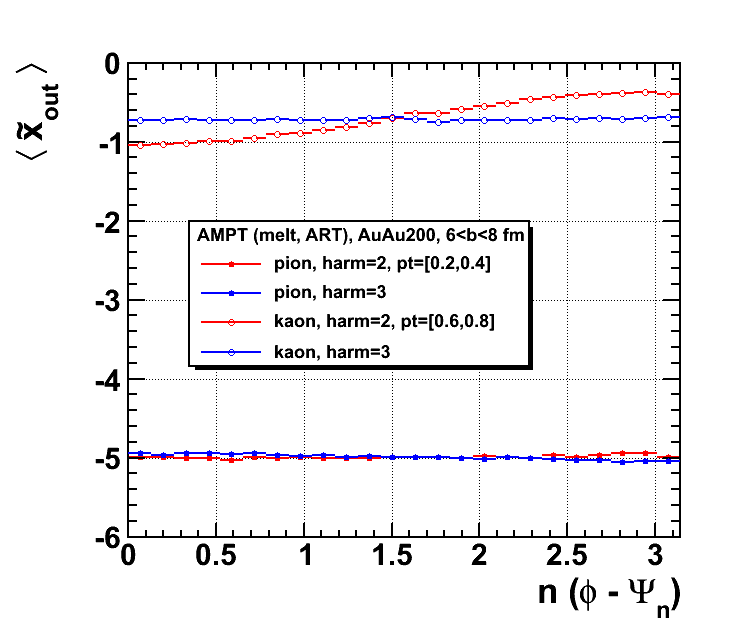}
\end{minipage} 
\caption{
AMPT model: the azimuthal dependence of the HBT radii (left and
middle), and radial shift in the production points of kaons and pions,
$\tilde{x}_{out}=x_{out}-V_t t$,
at $p_t\sim 1.5\, m$.
}
\label{fig:AMPTradii} 
\end{figure}

The AMPT model~\cite{Zhang:1999bd} has been extensively used in
studying the effect of the initial state fluctuations. We use this
model to study Au+Au collisions at
\snn=200~GeV. Figure~\ref{fig:AMPTradii} shows the HBT radii variation
as function of the azimuth relative to a given harmonic {\em event
plane} which includes the effects of finite reaction plane
resolution. Once again, the radii dependence on the azimuth is clearly
seen. A detailed investigation of the side-out relative phases,
dependece on the particle transverse momentum, etc. requires
additional investigations.

\vspace*{0.3 cm}
\noindent
\underline{\em Summary.} Using the blast wave and AMPT models we have
demostrated the possibility of femtoscopic higher
harmonics system shape analysis that promices new insights into the physics
of heavy ion collisions.

\end{document}